\documentclass[prl, superscriptaddress, showpacs, floatfix,
twocolumn]{revtex4}

\usepackage[]{graphicx}
\usepackage{amsmath}
\usepackage{amsfonts}

\def\ket#1{| #1 \rangle}

\def\kb#1#2{|#1\rangle\!\langle #2 |}
\def\II{1\!\mathrm{l}}
\def\A{{\cal A}}
\def\B{{\cal B}}
\def\C{{\cal C}}
\def\E{{\cal E}}
\def\H{{\cal H}}
\def\K{{\cal K}}
\def\M{{\cal M}}
\def\P{{\mathcal P}}
\def\R{{\cal R}}

\newcommand{\fA}{{\mathfrak{A}}}

\def\Tr{{\mathrm{Tr}}}

\newcommand{\qforal}{\quad\text{for all}\quad}

\newcommand{\fix}{{\rm Fix}\,}
\newtheorem{theorem}{Theorem}
\newtheorem{lemma}{Lemma}
\newtheorem{remark}{Remark}

\date{\today}

\begin{document}

\title{A Unified and Generalized Approach to Quantum Error Correction}

\author{David Kribs}
\affiliation{Institute for Quantum Computing, University of
Waterloo, ON Canada; and\\ Perimeter Institute for Theoretical
Physics, Waterloo, ON, Canada, N2L 2Y5} \affiliation{Department of
Mathematics and Statistics, University of Guelph, Guelph, ON,
Canada}
\author{Raymond Laflamme}
\affiliation{Institute for Quantum Computing, University of
Waterloo,  ON Canada; and\\ Perimeter Institute for Theoretical
Physics, Waterloo, ON, Canada, N2L 2Y5}
\author{David Poulin}
\affiliation{Institute for Quantum Computing, University of
Waterloo, ON Canada; and\\ Perimeter Institute for Theoretical
Physics, Waterloo, ON, Canada, N2L 2Y5}

\begin{abstract}
We present a unified approach to  quantum error correction, called
{\it operator quantum error correction}. This scheme relies on a
generalized notion of noiseless subsystems that is {\it not} restricted to the commutant of the interaction algebra. We arrive at the unified approach, which incorporates the
known techniques --- i.e. the standard error correction model, the
method of decoherence-free subspaces, and the noiseless subsystem
method --- as special cases, by combining active error correction
with this generalized noiseless subsystem method.  Moreover, we
demonstrate that the quantum error correction condition from the
standard model is a necessary condition for all known methods of
quantum error correction.
\end{abstract}

\maketitle

The possibility of protecting quantum information against
undesirable noise has been a major breakthrough for the field of
quantum computing, opening the path to potential practical
applications. In this paper, we show that the various techniques
used to protect quantum information all fall under the same
unified umbrella. First, we will review the standard model for
quantum error correction~\cite{Sho95a,Ste96a,BDSW96a,KL97a}, and
the passive error prevention methods of ``decoherence-free
subspaces''~\cite{PSE96,DG97c,ZR97c,LCW98a} and  ``noiseless
subsystems''~\cite{KLV00a,Zan01b,KBLW01a}. We shall then
demonstrate how the latter scheme admits a natural generalization,
and study the necessary and sufficient conditions leading to such
{\it generalized noiseless subsystems}. This generalized method in
turn motivates a unified approach --- called {\it operator quantum
error correction}
--- that incorporates all aforementioned techniques as special
cases. We describe this approach and discuss testable conditions
that characterize when error correction is possible given a noise model.
Moreover, we show that the standard error correction
condition is a prerequisite for any of the known forms of error
correction/prevention to be feasible.
\medskip

\noindent{\it The Standard Model} --- What could be called the
``standard model'' for quantum error correction~\cite{Sho95a,Ste96a,BDSW96a,KL97a}
consists of a triple $(\R,\E,\C)$ where $\C$ is a subspace, a {\it quantum
code}, of a Hilbert space $\H$ associated with a given quantum
system. The error $\E$ and recovery $\R$ are quantum operations on
$\B(\H)$, the set of operators on $\H$, such that $\R$ undoes the
effects of $\E$ on $\C$ in the following sense:
\begin{eqnarray}\label{reverse}
(\R\circ \E) \, (\sigma) = \sigma \qforal \sigma = P_\C \sigma P_\C,
\end{eqnarray}
where $P_\C$ is the projector of $\H$ onto $\C$. As a prelude to
what follows below, let us note that instead of focusing on the
subspace $\C$, we could just as easily work with the set of
operators $\B(\C)$ which act on $\C$.

When there exists such  an $\R$ for a given pair $\E,\C$, the
subspace $\C$ is said to be {\it correctable for $\E$}. The action
of the noise operation $\E$ can be described in an operator-sum
representation as $\E(\sigma) = \sum_a E_a \sigma E_a^\dagger$. While
this representation is not unique, all representations of a given
map $\E$ are linearly related: if $\E(\sigma) = \sum_b F_b \sigma
F_b^\dagger$, then there exists scalars $u_{ba}$ such that $F_b =
\sum_a u_{ba}E_a$. We shall identify the map $\E$ with any of its
error operators $\E = \{E_a\}$.  The existence of a recovery
operation $\R$ of $\E$ on $\C$ may be cleanly phrased in terms of
the $\{E_a\}$ as follows \cite{BDSW96a,KL97a}:
\begin{equation}
P_\C E_a^\dagger E_b P_\C = \lambda_{ab}P_\C \qforal a,b
\label{eq:standard}
\end{equation}
for some scalars $\lambda_{ab}$. Clearly, this condition is
independent of  the operator-sum representation of $\E$.
\medskip

\noindent{\it Noiseless Subsystems \& Decoherence-Free Subspaces}
--- Let $\E: \B(\H) \rightarrow \B(\H)$ be a quantum operation
with errors $\{E_a\}$. The algebra $\A$ generated by the set
$\{E_a,E_a^\dagger\}$ is a $\dagger$-algebra \cite{Dav96a}, called
the {\it interaction algebra}, and as such it is unitarily
equivalent to a direct sum of (possibly ``ampliated'') full matrix
algebras: $ \A \cong \bigoplus_J  \M_{m_J}\otimes \II_{n_J} . $

If $\E$ is a {\it unital} quantum operation, by which we mean that
the maximally mixed state $\II$ remains unaffected by $\E$ (i.e.,
$\E(\II)=\II$), then the fundamental {\it noiseless subsystem}
(NS) method of quantum error correction
\cite{KLV00a,Zan01b,KBLW01a} may be applied. This method makes use
of the operator algebra structure of the ``noise commutant'',
\[
\A' = \big\{\sigma\in\B(\H): E \sigma = \sigma E \,\,\,\forall
E\in\{E_a,E_a^\dagger\} \big\},
\]
to encode states that are immune to the errors of $\E$. As such,
it is in effect a method of error prevention. Notice that with the
structure of $\A$ given above, the noise commutant is unitarily
equivalent to $\A'\cong \oplus_J \II_{m_J} \otimes \M_{n_J}$.

In \cite{BS98b,Kri03a} it was proved that for unital $\E$, the
noise commutant coincides with the fixed point set for $\E$; i.e.,
\begin{eqnarray*}\label{fixedptthm}
\A' = \fix(\E) &=& \{ \sigma\in\B(\H): \E(\sigma)=\sigma\}.
\end{eqnarray*}
This is precisely the reason that $\A^\prime$ may be used to
produce NS for unital $\E$. We note that while many of the
physical noise models satisfy the unital constraint, there are
important non-unital models as well. Below we show how shifting
the focus from $\A^\prime$ to $\fix(\E)$ (and related sets) quite
naturally leads to a generalized notion of NS that applies to
non-unital quantum operations as well.

Note that the structure of $\A$ given above induces a natural
decomposition of the Hilbert space
\begin{equation*}
\H = \bigoplus_J \H^A_J\otimes\H^B_J,
\end{equation*}
where the ``noisy subsystems" $\H^A_J$ have dimension $m_J$ and  the ``noiseless subsystems" $\H^B_J$ have dimension $n_J$. For brevity,
we focus on the case where information is encoded in a single
noiseless sector of $\B(\H)$,  so
\begin{equation*}
\H = (\H^A \otimes H^B) \oplus \K \label{eq:decomp}
\end{equation*}
with $\dim(\H^A) = m$, $\dim(\H^B) = n$ and $\dim \K=\dim\H - mn$.
The generalization to multiple $J$'s is straightforward. We shall
write $\sigma^A$ for operators in $\B(\H^A)$ and $\sigma^B$ for
operators in $\B(\H^B)$. Thus the restriction of the noise
commutant $\A'$ to $\H^A\otimes\H^B$ consists of the operators of
the form $\sigma^{AB} = \II^A\otimes\sigma^B$ where $\II^A$ is the
identity element of $\B(\H^A)$. It is easy to see that such states
are immune to noise in the unital case.

For notational purposes, assume that ordered orthonormal bases
have been chosen for $\H^A =
{\mathrm{span}}\{\ket{\alpha_i}\}_{i=1}^m$ and $\H^B =
{\mathrm{span}}\{\ket{\beta_k}\}_{k=1}^n$ that yield the matrix
representation of the corresponding subalgebra of $\A'$ as $\II_m
\otimes \M_n$. We let
\begin{equation*}
\{ P_{kl} = \kb{\alpha_k}{\alpha_l}\otimes\II_{n}: 1\leq k,l \leq m\}
\end{equation*}
denote the corresponding family of ``matrix units'' associated with this
decomposition. In terms of these matrix units, the {\it minimal
reducing projectors} for $\A'$ are given by $P_k =
\kb{\alpha_k}{\alpha_k}\otimes\II_{n} = P_{kk}\in\A$. The
following equalities are readily verified and in fact are the
defining properties for a family of matrix units.
\begin{eqnarray*}
P_{kl} & =& P_k P_{kl} P_l  \quad \forall\, 1\leq k,l\leq m \\
P_{kl}^\dagger &=& P_{lk} \quad\quad\quad \forall\, 1\leq k,l \leq m \\
P_{kl}P_{l'k'} &=& \left\{ \begin{array}{cl} P_{kk'} & \mbox{if
$l=l'$} \\ 0 & \mbox{if $l\neq l'$} \end{array}\right..
\end{eqnarray*}
With these properties in hand, the following useful result may
be easily proved.

\begin{lemma}
The map $\Gamma = \{P_{kl}\}$ from $\B(\H)$ to itself satisfies the following two properties
\begin{eqnarray*}
\Gamma(\sigma) &=& \sum_{k,l}P_{kl}\sigma P_{kl}^\dagger \,\,\,\in \A' \\
\Gamma(\sigma^A\otimes\sigma^B) &\propto& \II^A\otimes \sigma^B  .
\end{eqnarray*}
for all operators $\sigma^A$, $\sigma^B$ and $\sigma \in \B(\H)$.
\label{lemma:map}
\end{lemma}

We note that the NS method contains the method of {\it
decoherence-free subspaces} (DFS)~\cite{DG97c,ZR97c,LCW98a} as a
special case. Specifically, if we are given an error operation
$\E$, then the DFS method encodes information in a subspace of the
system's Hilbert space that is immune to the evolution. However,
instead of working at the level of vectors, we could work at the
level of operators. In particular, as in the standard model, we
may identify a given Hilbert space $\H$ with the full algebra
$\B(\H)$ of operators acting on $\H$. In doing so, the DFS method
may be regarded as a special case of the NS method in the sense
that the DFS method in effect makes use of the ``unampliated''
summands, $ \II_{m_J}\otimes\M_{n_J}$ where $m_J=1$, inside the
noise commutant $\A^\prime$ for encoding information.
\medskip

\noindent{\it Generalized Noiseless Subsystems} --- We now describe a
generalized notion of noiseless subsystems that serves as a building block for the
unified approach to error correction discussed below and applies
equally well to {\it non-unital} maps. In the standard NS method,
the quantum information is encoded in $\sigma^B$; i.e., the state of
the noiseless subsystem. Hence, it is not necessary for the noisy
subsystem to remain in the maximally mixed state $\II^A$ under
$\E$, it could in principle get mapped to any other state.

In order to formalize this idea, define for a fixed decomposition
$\H = (\H^A\otimes\H^B) \oplus\K$ the set of operators
\[
\fA = \{\sigma\in\B(\H) : \sigma = \sigma^A\otimes\sigma^B,\,{\rm
for\,\,some}\,\, \sigma^A\,{\rm and}\, \sigma^B\}.
\]
Notice that this set has the structure of a semigroup and includes
operator algebras such as $\II^A\otimes\B(\H^B)$. For notational
purposes, we assume that bases have been chosen and define the
matrix units $P_{kl}$ as above, so that $P_k = P_{kk}$, $P_\fA =
P_1+\ldots + P_m$, $P_\fA\H = \H^A\otimes\H^B$, $P_\fA^\perp = \II
- P_\fA$ and $P_\fA^\perp\H = \K$. We also define a map $\P_\fA$
by the action $\P_\fA(\cdot) = P_\fA(\cdot) P_\fA$. The following
result leads to our generalized definition of NS.

\begin{lemma}
Given a fixed decomposition $\H = \H^A\otimes\H^B \oplus\K$ and a
map $\E$, the following three conditions are equivalent:
\begin{enumerate}
\item $\forall\sigma^A\ \forall\sigma^B,\ \exists \tau^A\ :\
\E(\sigma^A\otimes\sigma^B) = \tau^A\otimes\sigma^B$ \item $
\forall\sigma^B,\ \exists \tau^A\ :\ \E(\II^A\otimes\sigma^B) =
\tau^A \otimes \sigma^B$ \item $\forall\sigma\in \fA\ :\
\big(\Tr_A\circ \P_\fA\circ \E\big)(\sigma) =\Tr_A(\sigma)$.
\end{enumerate}
\label{lemma:generalNS}
\end{lemma}

\noindent{\it Proof.} The implications {\it 1.} $\Rightarrow$ {\it
2.} and {\it 1.} $\Rightarrow$ {\it 3.} are trivial. To prove {\it
2.} $\Rightarrow$ {\it 1.}, observe that $\sum_{k=1}^m
\kb{\alpha_k}{\alpha_k} = \II^A$, so condition {\it 2.}\@ implies
that for any $\ket\psi \in \H^B$,
\begin{equation}
\sum_{k=1}^m \E(\kb{\alpha_k}{\alpha_k}\otimes\kb\psi\psi) =
\tau^A\otimes\kb\psi\psi \label{eq:convex}
\end{equation}
for some $\tau^A \in \B(\H^A)$. Since $\E$ is a quantum
operation, $\sigma_{\psi,k} =
\E(\kb{\alpha_k}{\alpha_k}\otimes\kb\psi\psi)$ are positive for
$k=1,\ldots,m$. Equation~(\ref{eq:convex}) implies that
$\tau^A\otimes\kb\psi\psi$ is a convex combination of the
operators $\sigma_{\psi,k}$, which is only possible if
$\sigma_{\psi,k} =  \sigma^A_{\psi,k}\otimes\kb\psi\psi$ for some
positive $\sigma^A_{\psi,k}$. Through an application of the
Stinespring dilation theorem~\cite{Sti55a} and a linearity
argument, it follows that $\sigma^A_{\psi,k}$ does not depend on
$\psi$. Since the basis $\{\ket{\alpha_k}\}$ and the state
$\ket{\psi}$ were chosen arbitrarily, the result now follows from
the linearity of $\E$.

To prove {\it 3.} $\Rightarrow$ {\it 2.},  note that since $\E$
and $\Tr_B$ are trace preserving, {\it 3}.\@ implies that
$\big(\P_\fA \circ \E\big)(\sigma) = \E(\sigma)$ for all $\sigma \in
\fA$. By setting $\sigma = \II^A\otimes\kb\psi\psi$ as above, we
conclude from {\it 3}.\@ that $\E(\sigma) =
\tau^A\otimes\kb\psi\psi$ for some $\tau^A$. The rest follows
from linearity.\hfill$\square$

The subsystem $\H^B$ is said to be {\it noiseless}  when it
satisfies one --- and hence all --- of the conditions in
Lemma~\ref{lemma:generalNS}. It is clear from the third condition
that the fate of the noisy subsystem $\H^A$ has no importance:
only the information stored in the noiseless subsystem $\H^B$ must
be preserved by $\E$. Note that the generalized definition of NS
coincides with the standard definition when $dim(\H^A) = 1$.
Hence, the notion of DFS is not altered by this generalization.

Given this new notion of a NS, the crucial question is to
determine what are the necessary and sufficient conditions for a
map $\E = \{E_a\}$ to admit a NS described by a semigroup $\fA$.
Recall that the condition expressed by Eq.~(\ref{eq:standard})
gives an answer for standard error correction. The following
Theorem provides an answer to this question in the general
noiseless subsystem setting.

\begin{theorem}\label{thm:NS}
Let $\E = \{E_a\}$ be a quantum operation on $\B(\H)$ and let
$\fA$ be a semigroup in $\B(\H)$ as above. Then $\fA$ encodes a
noiseless subsystem (decoherence-free subspace in the case m=1)
--- as defined by any of the three conditions of
Lemma~\ref{lemma:generalNS} --- if and only if the  following two
conditions hold:
\begin{equation}
P_k E_a P_l = \lambda_{akl} P_{kl} \qforal a,k,l
\label{eq:cond1}
\end{equation}
for some set of scalars $\{\lambda_{akl}\}$ and
\begin{equation}
P_\fA^\perp E_a P_\fA = 0 \qforal a.
\label{eq:cond2}
\end{equation}
\end{theorem}

\noindent{\it Proof.} To prove the necessity of
Eqs.~(\ref{eq:cond1},\ref{eq:cond2}), note that
Lemma~\ref{lemma:map} and Lemma~\ref{lemma:generalNS} imply
\begin{equation}
\big(\Gamma\circ\E\circ\Gamma\big)(\sigma) \propto \Gamma(\sigma) \qforal \sigma
\in \B(\H).
\label{eq:conditionNS}
\end{equation}
By linearity, the proportionality factor cannot  depend on $\sigma$,
so the sets of operators $\{P_{ki}E_aP_{jl}\}$ and $\{\lambda
P_{k'l'}\}$ define the same map for some scalar $\lambda$. We may
thus find a set of scalars $\mu_{kiajl,k'l'}$ such that
\begin{equation}
P_{ki}E_aP_{jl} = \sum_{k'l'} \mu_{kiajl,k'l'} P_{k'l'}.
\end{equation}
Multiplying both sides of this equality on the  right by $P_l$ and
on the left by $P_k$, we see that $\mu_{ijakl,i'l'} = 0$ when
$k\neq k'$ or $l\neq l'$. This implies Eq.~(\ref{eq:cond1}) with
$\lambda_{akl} = \mu_{kkall,kl}$.

For the second condition, note that by definition $P_\fA^\perp
\sigma P_\fA^\perp = 0$ for all $\sigma \in \fA$. Together with
Lemma~\ref{lemma:map} and Lemma~\ref{lemma:generalNS}, this
implies $P_\fA^\perp \E(\Gamma(\sigma)) P_\fA^\perp = 0$ for all
$\sigma \in \B(\H)$. Equation~(\ref{eq:cond2}) follows from this
observation via a consideration of the operator-sum representation
for $\E$.

To  prove sufficiency, we use the definitions $\II = P_\fA +
P_\fA^\perp$ and $P_\fA = \sum_{k=1}^m P_k$ to establish for all
$\sigma\in\fA$
\begin{eqnarray*}
\E(\sigma) &=& (P_\fA + P_\fA^\perp) \sum_a E_a\sigma E^\dagger_a(P_\fA + P_\fA^\perp) \\
&=&\sum_a P_\fA E_a \sigma E^\dagger_a P_\fA \\
&=& \sum_{a,k,k'} P_kE_a\sigma E_a^\dagger P_{k'}.
\end{eqnarray*}
Combining this with the identity $\sigma^A\otimes\sigma^B= P_\fA( \sigma^A\otimes\sigma^B) P_\fA =
\sum_{l,l'} P_l (\sigma^A \otimes \sigma^B) P_{l'}$ implies
\begin{eqnarray*}
\E(\sigma^A\otimes\sigma^B) &=& \sum_{a,k,k',l,l'} P_kE_aP_l(\sigma^A\otimes\sigma^B) P_{l'} E^\dagger_a P_{k'} \\
&=& \sum_{a,k,k',l,l'} \lambda_{akl}\overline{\lambda}_{ak'l'}
P_{kl} (\sigma^A\otimes\sigma^B) P_{l'k'}.
\end{eqnarray*}
The proof now follows from the fact that the matrix units $P_{kl}$
act trivially on the $\B(\H^B)$ sector. \hfill$\square$

Conditions Eqs.~(\ref{eq:cond1},\ref{eq:cond2}) do not necessarily
imply that the noiseless operators are in the commutant of the
interaction algebra $\A = \{E_a\}$ since $P_\fA E_a P_\fA^\perp$
is not necessarily equal to zero. Hence, this generalization does
indeed admit new possibilities.
\medskip

\noindent{\it The Unified Approach} --- The unified scheme for
quantum error correction consists of a triple $(\R,\E,\fA)$ where
again $\R$ and $\E$ are quantum operations on some $\B(\H)$, but
now $\fA$ is a semigroup in $\B(\H)$ defined as above with respect
to a fixed decomposition $\H = (\H^A \otimes \H^B) \oplus \K$.
Given such a triple $(\R,\E,\fA)$ we say that $\fA$ is {\it
correctable for $\E$} if
\begin{eqnarray}\label{newid}
\big(\Tr_A \circ \P_\fA \circ\R \circ \E \big) (\sigma) =
\Tr_A(\sigma) \qforal \sigma \in \fA.
\end{eqnarray}

In other words, $(\R,\E,\fA)$ is a correctable triple if the
$\H^B$ sector of the semigroup $\fA$ encodes a noiseless subsystem
of the error map $\R\circ\E$. Thus, substituting $\E$ by
$\R\circ\E$ in Lemma~\ref{lemma:generalNS} offers  alternative
equivalent definitions of a correctable triple. Observe that the
standard model for error correction is given by the particular
case in this model that occurs when $m=1$.
Lemma~\ref{lemma:generalNS} shows that the generalized (and
standard) NS and DFS methods are captured in this model when $\R =
{\rm id}$ is the identity channel and, respectively, $m\geq 1$ and
$m=1$.

We next present  a mathematical condition that characterizes
correctable codes for a given channel $\E$ in terms of its error
operators and generalizes Eq.~(\ref{eq:standard}) for the standard
model. Again, we assume that matrix units $P_{kl}$ associated with
the noise commutant have been defined as above.

\begin{theorem}\label{unifiedthm}
Let $\E = \{E_a\}$ be a quantum operation on $\B(\H)$ and let $\fA$ be a semigroup in $\B(\H)$ as above. If there is
a quantum operation $\R$ on $\B(\H)$ such that
\begin{eqnarray}\label{correctopn}
\big( \Tr_A\circ\P_\fA\circ\R \circ \E \big) (\sigma) = \Tr_A(\sigma)
\qforal \sigma \in \fA,
\end{eqnarray}
then there are scalars $\Lambda = \{ \lambda_{abkl} \}$ such that
\begin{eqnarray}\label{condition}
P_{k} E_a^\dagger E_b P_l = \lambda_{abkl} P_{kl} \qforal
a,b,k,l.
\end{eqnarray}
\end{theorem}

\noindent{\it Proof.} As noted above $(\R,\E,\fA)$ being a
correctable triple implies that $\fA$ encodes a generalized
noiseless subsystem of the map $\R\circ\E$. Applying
Theorem~\ref{thm:NS}, and in particular condition
Eq.~(\ref{eq:cond1}), to the map $\R\circ\E$ implies the existence
of a set of scalars $\mu_{cakl}$ for which $P_kR_cE_aP_l =
\mu_{cakl}P_{kl}$. It now follows from Eq.~(\ref{eq:cond2})
applied to the map $\R\circ\E$ and $P_\fA = \sum_j P_j$ that
\begin{eqnarray*}
P_kE_a^\dagger E_b P_l
&=& \sum_c P_{k}^\dagger E_a^\dagger R_c^\dagger R_c E_b P_{l} \\
&=& \sum_{c,j} P_{k}^\dagger E_a^\dagger R_c^\dagger P_j^\dagger P_j R_c E_b P_{l} \\
&=& \sum_{c,j} \overline{\mu}_{cajk}\mu_{cbjl}P_{jk}^\dagger P_{jl} \\
&=& \left(\sum_{c,j}  \overline{\mu}_{cajk}\mu_{cbjl}
\right) P_{kl},
\end{eqnarray*}
and this completes the proof of the Theorem. \hfill$\square$

\begin{remark}
The condition Eq.~(\ref{condition}) is independent of the choice
of basis $\{\ket{\alpha_i}\}$ that defines the family $P_{kl}$ and
of the operator-sum representation of $\E$. In particular, under
the changes $\ket{\alpha'_k} = \sum_l u_{kl}\ket{\alpha_l}$ and
$F_a = \sum_b w_{ab} E_b$, the scalars $\Lambda$ change to
$\lambda_{abkl}' = \sum_{a'b'k'l'}
\overline{u}_{kk'}u_{l'l}\overline{w}_{aa'}w_{bb'}
\lambda_{abkl}$.
\end{remark}

Equation~(\ref{condition}) generalizes the quantum error
correction  condition Eq.~(\ref{eq:standard}) to the case where
information is encoded in operators, not necessarily restricted to
act on a fixed code subspace $\C$. However, observe that setting
$k= l$ in Eq.~(\ref{condition}) gives the standard error
correction condition Eq.~(\ref{eq:standard}) with $P_\C = P_k$.
This leads to the following result.

\begin{theorem}
If $(\R,\E,\fA)$ is a correctable triple for some semigroup $\fA$
defined as above, then $(\P_k\circ\R,\E,P_k\fA P_k)$ is a
correctable triple according to the standard definition
Eq.~(\ref{eq:standard}), where $P_k$ is any minimal reducing
projector of $\fA$, and the map $\P_k$ is defined by $\P_k(\cdot)
= \sum_l P_{kl}(\cdot )P_{kl}^\dagger$. \label{thm:standard}
\end{theorem}

\noindent{\it Proof.} The error correction condition
Eq.~(\ref{newid}) and Lemma~\ref{lemma:generalNS} imply that for
all $\sigma^B$ there is a $\tau^A$ such that
\begin{equation*}
\big(\R\circ\E\big)\big(P_k (\II^A \otimes \sigma^B) P_k\big)
\propto \tau^A\otimes\sigma^B.
\end{equation*}
Observe that $\P_k(\tau^A\otimes\sigma^B)
\propto \kb{\alpha_k}{\alpha_k}\otimes\sigma^B$ for all $\sigma^B$ and
$\tau^A$. Combining these two observations, we conclude that
\begin{equation*}
\big(\P_k\circ\R\circ\E\big)\big(P_k(\II^A\otimes\sigma^B)P_k\big)
\propto P_k(\II^A\otimes\sigma^B)P_k,
\end{equation*}
completing the proof. \hfill$\square$

Theorem~\ref{thm:standard} has important consequences. Given a map
$\E$, the existence of a correctable code subspace $\C$ --- captured by
the standard error correction condition Eq.~(\ref{eq:standard})
--- is a prerequisite to the existence of any known type of error
correction/prevention scheme (including the generalizations
introduced in the present paper). Moreover,
Theorem~\ref{thm:standard} shows how to transform any one of these
error correction/prevention techniques into a standard error
correction scheme.

Finally, note that Theorem~\ref{unifiedthm} sets  necessary conditions for the
possibility of operator quantum error corrections, but does not
address sufficiency. At the time of writing, we have not proved
sufficiency in full generality. We have, however, demonstrated
that these conditions are sufficient for a number of motivating
special cases. This topic will be discussed in an upcoming paper~\cite{KLLP05a}.
\medskip

\noindent{\it Conclusion} --- We have presented a general model for quantum
error correction, called {\it operator quantum error correction},
that unifies the fundamental paradigms. In doing so, we have
generalized the method of {\it active} error correction by
implementing the condition at the level of operators rather than
subspaces. We have also generalized the notion of noiseless
subsystems by relaxing the constraints imposed on the ``noisy"
sector of the algebra; i.e., that it remains in the maximally
mixed state.  In addition, we have demonstrated that the standard
error condition Eq.~(\ref{eq:standard}) is a necessary condition
for any type of error correction --- either passive or active ---
to be possible, and we have shown how to convert any such scheme
into a standard error correction protocol.

We thank Maia Lesosky, Harold Ollivier, Rob Spekkens and our other
colleagues at IQC and PI for helpful discussions. This work was
supported in part by funding from NSERC, CIAR, MITACS, NATEQ, and ARDA.


\end{document}